\documentclass[a4paper]{article}

\usepackage{amsfonts}
\usepackage{amssymb}
\usepackage[margin=2cm]{geometry}
\usepackage[dvips]{graphicx}

\usepackage[T1]{fontenc}
\usepackage[latin2]{inputenc}

\usepackage{amsmath}
\usepackage{bbm}

\begin{document}


\title{Quantum dynamical entropy and decoherence rate}

\author{Robert Alicki$^{1}$, Artur {\L}ozi{\'n}ski$^2$,
 Prot Pako{\'n}ski$^{2}$, and Karol {\.Z}yczkowski$^{2,3}$ \\ 
  {\small
$^1$Institute of Theoretical Physics and Astrophysics, University
of Gda\'nsk,  Wita Stwosza 57, PL 80-952 Gda\'nsk, Poland}\\
{\small $^2$Institute of Physics, Jagiellonian University,
ul Reymonta 4, 30-059 Krak{\'o}w, Poland} \\
{\small $^3$Center for Theoretical Physics, Polish Academy of Sciences,
   al. Lotnik{\'o}w 32/44, 02-668 Warszawa, Poland}
}   

\date{\today}
\maketitle

\begin{abstract}
We investigate quantum dynamical systems defined on a finite dimensional Hilbert space
and subjected to an interaction with an environment.  The rate of decoherence of
initially pure states, measured by the increase of their von Neumann entropy, averaged
over an ensemble of random pure states, is proved to be bounded from above by the
partial entropy used to define the  ALF dynamical entropy.  The rate of decoherence
induced by the sequence of the von Neumann projectors measurements is shown to be
maximal, if the measurements are performed in a randomly chosen basis.  The
numerically observed linear increase of entropies is attributed to free-independence
of the measured observable and the unitary dynamical map.
\end{abstract}

{\small \hspace{2.6mm} PACS numbers: 03.65.Yz, 05.45.Mt}

{\small \hspace{2.6mm} Keywords: quantum chaos, entropy production, decoherence}

\section{Introduction}
The notion of chaos in classical mechanics is well established,
and any dynamical system characterized by positive Kolmogorov--Sinai
entropy is called {\sl chaotic} \cite{KH96}. On the other hand it is not
at all easy to generalize the definition of chaos for quantum theory 
\cite{OP93,AF01}. There exist numerous attempts to define the quantum counterpart of
the Kolmogorov-Sinai entropy both for finite and infinite quantum systems (see
\cite{SZ94} and references therein).  However, only  two of them: CNT--entropy
\cite{CNT87} and ALF-entropy \cite{AF94} provide nonequivalent notions of quantum
dynamical entropy which satisfy the following conditions:

1) can be formulated in an abstract algebraic framework valid for general commutative
(classical) and noncommutative (quantum) dynamical systems,

2) coincide with the Kolmogorov-Sinai entropy when applied to classical systems,

3) can be rigorously computed for several examples on noncommutative dynamical systems
(different types of quantum shifts, automorphisms of quantum tori, fermionic
quasi-free systems).

In contrast to the coherent-states quantum entropy defined in \cite{SZ94,SZ99}, both
CNT and ALF entropies are always equal to zero for quantum systems with
finite-dimensional Hilbert spaces. In particular this concerns the systems being
quantizations of classically chaotic dynamical systems defined on a compact 
phase-space, we are  going to analyze in this work.  For the case of  ALF entropy we
can easily understand the mechanism leading to the apparent lack of "correspondence
principle" for the K-S entropy. In fact we can see that classical-quantum
correspondence exists provided the proper order of limit procedures is used.
Recently this problem has been studied rigorously for both CNT and ALF entropies in
the case of quantized Arnold cat maps \cite{BCCFV03}.

A typical quantum state coupled with an environment suffers decoherence, i.e. a
generic pure state becomes mixed as a result of the non-unitary dynamics.  As shown
by Zurek and Paz \cite{zurek94}, the initial rate of decoherence is governed by the
classical dynamical entropy $h$. Vaguely speaking, any classical density evolving in a
2-dimensional phase space of a discrete invertible chaotic map $T$, is squeezed along
the stable manifold and simultaneously stretched along the unstable manifold with the
rate determined by the classical Lyapunov exponent $\lambda$. In a similar way the
corresponding quantum wave packet is stretched. So in a generic case, it becomes
coupled with exponentially increasing number of states localized in the phase space. A
natural assumption that these states are distinguishable by the environment implies
the initially linear growth of the von~Neumann entropy of a typical pure state, with
the slope given by $h[T]$.

Detailed investigation of the rate of decoherence in various setups is a subject of a
considerable recent interest \cite{MP00,BPS02,MSS03,BCZ03}.  The main aim of this work
is to find a more precise bound for the increase of the average von Neumann entropy in
time.  We analyze the decoherence in finite quantum chaotic systems subjected to the
sequence of periodical measurement process and establish a link between the rate of
decoherence and the partial entropy used for the definition of the ALF--entropy.

The paper is organized as follows. In section \ref{sec_ALF} we recall the definition
of the ALF quantum dynamical entropy.  In section \ref{sec_cllim} we discuss
semiclassical limit of quantum maps and in section \ref{sec_correspond} analyze
classical limit of the ALF entropy.  In section \ref{sec_entprod} we analyze the rate
of decoherence and provide the upper bound (\ref{eq_mainineqaulity}) for the time
evolution of the mean von Neumann entropy, averaged over the set of random initial
pure states.  These general results are used in section \ref{sec_decoh} by studying
the decoherence in a model system: periodically measured quantum baker map. Discussion
of the results obtained in the context of the free-independent variables is provided
in chapter \ref{sec_free_indep}.

\section {ALF-dynamical entropy}
\label{sec_ALF}

We describe an abstract, discrete-time reversible dynamical system in terms of the
Hilbert space ${\cal H}$, the unitary dynamical map $U$ and the normalized vector
$|\Omega \rangle \in {\cal H}$.  The relevant (generally complex) physical
observables form a $*$-algebra ${\cal A}$ (i.e. a linear space with multiplication and
adjoint operation) of bounded operators acting on ${\cal H}$ closed under the dynamics
governed by $U$, i.e ${\cal A}\ni X\mapsto UXU^{-1} \equiv {\cal U}(X)\in {\cal A}$.
We assume also that the state given by $|\Omega\rangle{}$ restricted to the algebra
${\cal A}$ is time invariant, i.e. $\langle{}\Omega|{\cal U}(X)|\Omega\rangle{} =
\langle{}\Omega|X|\Omega\rangle{}$.  In particular for the classical system $(\Gamma,
T, d\gamma)$ with the reduced phase-space $\Gamma$, dynamical map $\gamma\mapsto
T(\gamma)$, and the invariant probability measure $d\gamma$, we obtain the Koopman's
formalism with ${\cal H} = L^2(\Gamma, d\gamma)$, $(U\psi)(\gamma) = \psi
(T(\gamma))$,  $|\Omega\rangle{} = ({\rm function}\equiv 1)$ and ${\cal A}$ the
algebra  of all measurable bounded functions on the phase-space $\Gamma$ treated as
multiplication operators on $L^2(\Gamma, d\gamma)$. For a finite quantum mechanical
principal system $S$ the Hilbert space ${\cal H}= {\cal H}_S\otimes {\cal H}_A$
describes the composite "system + ancilla" $(S+A)$ with  ${\rm dim} {\cal H}_S = {\rm
dim} {\cal H}_A = d$, and the algebra ${\cal A}$ consists of elements of the product
form $X = X_S\otimes{\mathbbm 1}_A $.  The state  $|\Omega\rangle{}$ is the
``purification'' of a generally mixed time invariant reference state of $S$. We have
certain freedom in the choice of operator $U$. The simplest one is $U = U_S\otimes
{\mathbbm 1}$, but it is sometimes convenient to put $U = U_S\otimes {\hat U}_S$,
where ${\hat U}_S$ is a properly defined ``transposition'' of $U_S$ such that
$U|\Omega\rangle = |\Omega\rangle $.  This second choice is natural in the theory of
infinite quantum dynamical systems usually formulated in terms of $C^*$-algebras for
which the present framework corresponds to the GNS-representation. 

ALF-entropy is defined in several steps using the notion of {\em partition of
unity} ${\bf X} = \{X_1, X_2,\dots ,X_k;\; X_j\in~{\cal A},\linebreak \sum _{j=1}^k
X_j^{\dagger}X_j = {\mathbbm 1}\}$. The partition of ${\cal H}_S$ generates a
corresponding tensor-product partition of the composed space $\cal H$, namely ${\bf
X}\otimes{\mathbbm 1} = \{X_1\otimes{\mathbbm 1},\dots ,X_k\otimes{\mathbbm
1}\}$. To shorten the notation we will write in the sequel $\bf X$ instead of ${\bf
X}\otimes{\mathbbm 1}$. Partitions of unity can be composed, ${\bf
X}\circ{\bf Y} = \{X_j Y_m \}$ and evolved ${\cal U}({\bf X}) = \{UX_jU^{\dagger}\}$
to produce finer partitions, ${\bf X}^t = {\cal U}^{t-1}({\bf X})\circ\cdots\circ{\cal
U}({\bf X})\circ{\bf X}$.  We use the following notation for the multi-time
correlation matrices given by 
\begin{eqnarray}
\sigma[{\bf X}^t]_{i_1,\dots ,i_t;j_1,\dots ,j_t}  = 
\langle{}\Omega |X^{\dagger}_{j_1}{\cal U}(X^{\dagger}_{j_2})\dots {\cal
U}^{t-1}(X^{\dagger}_{j_t}) {\cal U}^{t-1}(X_{i_t})\dots {\cal U}
(X_{i_2})X_{i_1}\Omega\rangle{}\nonumber\\ 
=\langle{}\Omega |X^{\dagger}_{j_1}U^{\dagger}
X^{\dagger}_{j_2}U^{\dagger}\dots X^{\dagger}_{j_t} X_{i_t}\dots U X_{i_2}U
X_{i_1}\Omega\rangle{}\ \text{.}
\end{eqnarray}
Here $\sigma[{\bf X}^t]$ is a positively defined, $k^t \times k^t$ complex-valued
matrix with a trace equals one. By $S_t[{\bf X}, U]$ we denote its von Neumann entropy 
\begin{equation}
S_t[{\bf X},U] = -{\rm tr} \bigl(\sigma[{\bf X}^t]\ln \sigma[{\bf X}^t]\bigr)
= -{\rm tr} \bigl(\Omega[{\bf X}^t]\ln \Omega[{\bf X}^t]\bigr) \text{,}
\label{eq_st}
\end{equation}
where $ \Omega[{\bf X}^t]$ is a finite range density operator acting on ${\cal H}$ 
\begin{equation}
\Omega[{\bf X}^t] = \sum_{j_1,j_2,\dots j_t=1}^k |U X_{j_t}\dots U X_{j_2}U
X_{j_1}\Omega\rangle{}\langle{}U X_{j_t}\dots U X_{j_2}U X_{j_1}\Omega| \ \text{.}
\end{equation}
Equality (\ref{eq_st}) follows from the fact that the spectrum of the operator
$\sum_{j=1}^k |j\rangle{}\langle{}j|$ is identical (including degeneracies) to the
spectrum of the $k\times k $ matrix $[\langle{}i |j\rangle{}]$ except the eigenvalues
equal to zero.

For any partition of unity ${\bf X}$, one may introduce the corresponding dynamical
map $\Phi_{\bf X}$ in the Schr\"odinger picture
\begin{equation}
\rho\ \mapsto\  \Phi_{\bf X}(\rho) = \sum_{j=1}^k X_j \rho X_j^{\dagger} \text{.}
\label{eq_supero}
\end{equation}
This map sends an arbitrary density operator $\rho$ in the set of density operators.
Iterating the state $|\Omega\rangle\langle\Omega|$ $t$-times by the map $\Psi_{U{\bf
X}}$ we obtain
\begin{equation}
\Omega[{\bf X}^t] = [\Phi_{U\bf X}]^t(|\Omega\rangle{}\langle{}\Omega|)\ \text{,}
\label{eq_Omega}
\end{equation}
where $U{\bf X} = \{UX_1,UX_2,\dots ,UX_k\}$.

The formula of above suggests a new interpretation of $S_t[{\bf X},U]$ as the entropy
of the density matrix obtained by repeated measurements performed on the evolving
system plus ancilla with the initial pure entangled state~$|\Omega\rangle{}$. 

The dynamical entropy of partition ${\rm h}[{\bf X},U]$ is defined as a limit
\begin{equation}
{\rm h}[{\bf X},U]  = \limsup_{t\to\infty}\frac{1}{t} S_t[{\bf X},U] \ \text{.}
\label{eq_limtinfty}
\end{equation}
Finally, the dynamical entropy of $U$ is a supremum over a given class $\{\bf X\}$ of
physically admissible partitions \cite{AF94}
\begin{equation}
{\rm h}[U] = \sup_{\{\bf X\}}{\rm h}[{\bf X},U]\ \text{,} 
\label{eq_qedyn} 
\end{equation}
i.e. over all generalized measurement processes.  One can choose as $\{\bf X\}$ all
partitions in the algebra ${\cal A}$ or restrict the supremum to bistochastic ones
(satisfying $\sum X_j X_j^{\dagger} = {\mathbbm 1}$), or even partitions with $X_j$
proportional to unitary $U_j$, namely $X_j = p_jU_j,\; p_j \geq 0,\; \sum_{j}p_j = 1$.
Although in general one obtains different types of ALF-entropies they all coincide in
the classical case and for all known examples of noncommutative dynamical systems.

Since for any finite quantum system the entropy $S_t[{\mathbf X},U]$ is limited
\begin{equation}
S_t[{\bf X},U] \leq {\text{min}}\{t\ln k , d\}\ ,
\label{eq_stmin}
\end{equation}
the asymptotic rate (\ref{eq_limtinfty}) of the entropy production gives zero,
independently of the investigated unitary dynamics $U$. On the other hand, one may
analyze the initial rate of the entropy $S_t$, which at small times was shown
\cite{AMM96} to be determined by the classical entropy $h$.
\section{Classical limit of quantum maps}
\label{sec_cllim}

Consider a classical dynamical system with a compact (reduced) phase-space $\Gamma$
equipped with the probability measure $d\gamma$. This measure is assumed to be
invariant with respect to the dynamical map $\gamma\mapsto T(\gamma)$. We say that
this system $(\Gamma, T, d\gamma)$ is a classical limit of the sequence of quantum
systems if:

a) there exists a sequence $({\bf C}^{d(n)} , U_n ; n= 1,2,\dots )$ of $d(n)$- dimensional
Hilbert spaces and $d\times d$- unitary matrices,

b) there exist a {\em quantization procedure} which to any real function $f(\gamma)$
(usually satisfying some smoothness properties) associates the sequence of self adjoint
operators $F^{(n)}$ acting on ${\bf C}^{d}$.

c) for any set of observables $f_1 , f_2 ,\dots ,f_k$ and any sequence of time steps $t_1
, t_2 ,\dots ,t_k$ the correlation functions converge,
\begin{equation}
\lim_{n\to\infty} \frac{1}{d(n)}{\rm Tr}\bigl(U_n^{t_1}F_1^{(n)}U_n^{-t_1} 
U_n^{t_2}F_2^{(n)}U_n^{-t_2}\cdots U_n^{t_k}F_k^{(n)}U_n^{-t_k}\bigr)=
\int_{\Gamma} d\gamma f_1(T^{t_1}(\gamma))f_2(T^{t_2}(\gamma))\cdots
f_k(T^{t_k}(\gamma)) \ .
\label{eq_cllim}
\end{equation}
Both, the quantization procedure $f_j\mapsto F_j$ and the choice of $U_n$ are not unique.
The maximally mixed states $\rho_{\star} = {\mathbbm 1}/d$ correspond to the
uniform normalized measure $d\gamma$.

The example of such structure has been rigorously studied for the Arnold cat maps in
the recent paper \cite{BCCFV03}.

To present an example of a family of quantum maps, corresponding to certain classical
system, we are going to recall the construction of quantum baker map, originally due
to Balazs and Voros \cite{balazs89} and modified later in
\cite{saraceno90,saraceno92,bievre96}.  Classical baker map is defined as a
transformation of a unit square -- the compact phase space $\Gamma$ with the
coordinates $q$ (position) and $p$ (momentum),
\begin{equation} \label{clasmap}
\Gamma \ni \gamma = (q,p) \rightarrow T_B(\gamma) =
\left( 2q-[2q], (p+[2q])/2 \right) \in \Gamma,
\end{equation}%
where $[2q]$ denotes the integer part of $2q$. This map is hyperbolic and its
Kolmogorov-Sinai entropy is equal to $\ln 2$. Such a transformation may be quantized
in a finite Hilbert space ${\bf C}^{d}$. In an ordered orthonormal basis named
position eigenbasis $\{e_j\}$ we introduce a periodic translation operator
\begin{equation} \label{poseigv}
U e_j = e_{j+1}\ , \quad j = 1, \ldots, d-1 \ , \ U e_{d} = e_1  \ .
\end{equation}
Diagonalization of $U$ leads to the conjugated basis -- momentum eigenbasis $\{\tilde
e_k\}$,
\begin{equation} \label{momeigv}
U \tilde e_k = \exp(2\pi i k/d) \tilde e_k \ .
\end{equation}
Analogously to~(\ref{poseigv}) the translation operator in the momentum
eigenbasis is introduced,
\begin{equation}
V \tilde e_k = \tilde e_{k+1} \ , \ V \tilde e_{d} = \tilde e_1
\ , \quad k = 1, \ldots, d-1 \ .
\end{equation}
This operator is diagonal in the position eigenbasis,
\begin{equation}
V e_j = \exp(-2\pi i j/d) e_j \ ,
\end{equation}
and the transformation between position and momentum basis is given by the discrete
Fourier transform ${\cal F}_{d}$,
\begin{equation}
\tilde e_k = \sum_j \left[ {\cal F}_{d} \right]_{kj} e_j =
\sum_j \frac{1}{\sqrt{d}} \; e^{-2\pi i k j/d} \; e_j \ .
\end{equation}
Having defined the group of translation operators corresponding to a classical torus
it is possible~\cite{balazs89,saraceno90,saraceno92} to link the unitary operator
\begin{equation} \label{quantmap}
U_B = \left( {\cal F}_{d} \right)^{-1} \cdot \left( \begin{array}{cc}
{\cal F}_{d/2} & 0 \\ 0 & {\cal F}_{d/2} \end{array} \right) \ ,
\end{equation}
acting on ${\bf C}^{d}$, where $d$ is even integer (e.g. $d=2n$), to the
classical transformation defined by (\ref{clasmap}). The translation operators allow
one also to define a finite dimensional operator corresponding to any classical
observable described by a continuous function $f$ on $\Gamma$. Let us defined the
Fourier expansion of $f$,
\begin{equation}
f(q,p) = \sum_{j,k} a_{jk} \; e^{-2\pi ijq/d} \; e^{2\pi ikp/d} \ .
\end{equation}
Thus the operator $F^{(n)}$ corresponding to observable $f$ may read as follows
\begin{equation}
F^{(n)} = \sum_{j,k} a_{jk} \; V^j \; U^k \ .
\end{equation}
As it was mentioned above the quantization procedure is not unique.  Another set of
operators $F^{(n)}$ may be obtained if we use different ordering of translation
operators, since they do not commute, $UV = VU e^{2\pi i/d}$. It is also possible to
generalize whole quantization procedure by introducing translation operators which are
not exactly periodic, but periodic up to a phase factor (e.g. $U^{d} = e^{2\pi i
\chi_p/d} \, {\mathbbm 1}$, $V^{d} = e^{2\pi i \chi_q/d} \,{\mathbbm 1}$)
\cite{saraceno90,bievre96}.

Different properties of such a quantum baker map were studied in \cite{balazs89, 
saraceno90,saraceno92,bievre96}, and the correspondence with the classical system
(\ref{clasmap}) was established. Although we cannot provide a formal proof
 that the quantization (\ref{quantmap}) satisfies the property (\ref{eq_cllim}),
 we are going to use this model in further numerical investigations.

\section{Correspondence principle for dynamical entropy}
\label{sec_correspond}

For any finite dimensional quantum system inequality (\ref{eq_stmin}) holds, so the
quantum dynamical entropy $h[U] = 0$. This fact is sometimes interpreted as the lack
of correspondence principle for dynamical entropy. In section \ref{sec_cllim} we
defined a family of quantum maps, parametrized by an integer index $n$, such that in
the semiclassical limit $n\mapsto \infty$ the dimension $d(n)$ becomes infinite.
Taking a sequence of quantum systems $ ({\bf C}^{d} , U_n )$ with the classical limit
$(\Gamma , T, d\gamma)$ in the sense defined above (\ref{eq_cllim}), we may start with
a functional partition of unity ${\bf f}=\{f_1,f_2,\dots ,f_k; \sum_j |f_j(\gamma)|^2
=1\}$ and construct its quantum counterparts ${\bf F}_n=
\{F^{(n)}_1,F^{(n)}_2,\dots ,F^{(n)}_k\}$ using a suitable quantization procedure. The
entropy $S_t[{\bf F}_n, U_n]$ is computed using Eq. (\ref{eq_st}),
\begin{equation}
\Omega[{\bf F}_n^t] = 
[\Phi_{U_n{\bf F}_n}]^t\otimes {\mathbbm 1}(|\Psi_n\rangle{}\langle{}\Psi_n|)\ \text{,}
\label{eq_qcorfun}
\end{equation}
where 
\begin{equation}
|\Psi_n\rangle{} = \frac{1}{\sqrt{d}}\sum
_{m=1}^{d}|e_m\rangle{}\otimes|e'_m\rangle{}\ ,\ \ \ \ \{|e_m\rangle{}\},
\{|e'_m\rangle{}\}-\ {\rm basis\ in}\ {\bf C}^{d}
\end{equation}
is the purification of the tracial state ${\mathbbm 1}/d$ of the system given in
terms of the maximally entangled vector in ${\bf C}^{d}\otimes {\bf C}^{d}$.  Then
according to the Eq. (\ref{eq_cllim})
\begin{equation}
S_t[{\bf f},T] = \lim_{n\to\infty} S_t[{\bf f}_n, U_n]\ ,
\end{equation}
the classical dynamical entropy of the partition can be recovered by taking first the
classical limit $n\to\infty$ and then long time limit $t\to\infty$. Usually, for a
given classical system with the K-S entropy ${\rm h}[T]$ there exist many  "optimal
partitions"  ${\bf f}$ with $\ln k \geq {\rm h}[T]$ for which $S_t[{\bf f},T] \approx
t\cdot {\rm h}[T]$ with a given accuracy or even exactly ( generating partitions,
Markovian partitions \cite{AF94}). Therefore we can expect that for large enough $n$
and the optimal choice of the partition ${\bf f}$ the entropy $S_t[{\bf F}_n, U_n]$
displays linear growth with the rate given by the K-S entropy ${\rm h}[T]$ for $t $
below $t_{max}= 2\ln d/{\rm h}[T]$ and then saturates at the maximal value $2\ln
d$. For the regular dynamics $T$ with ${\rm h}[T] =0$ we expect a slower
(logarithmic) increase of $S_t[{\bf F}_n, U_n]$ up to the maximal value.

In formula (\ref{eq_qcorfun}) any purification of the tracial state can be used while
for concrete computations some choices could be better than the others. The set ${\cal
E}$ of pure states $\{|\alpha\rangle : \langle \alpha|\alpha \rangle =1 \}$ in
${\mathbb C}^d$ equipped with the probability measure $d\alpha$ and satisfying
\begin{equation}
\int_{\cal E} d\alpha\,|\alpha\rangle \langle \alpha| = \frac{1}{d} {\mathbbm 1}\ ,
\end{equation}
will be called {\em complete set of vectors}.  The natural examples of ${\cal E}$ are
orthonormal basis or coherent states generated by the irreducible representations of
certain compact Lie groups on ${\bf C}^d$. The distinguished example is the set of all
pure states ${\cal P}_d={\mathbbm C}P^{d-1}$%
with  the natural unitary invariant probability measure (Fubini--Study measure).  The
following theorem provides the most general representation of  maximally entangled
vectors in terms of complete sets.

{\bf Theorem 1} 

1) Any purification of the tracial state ${\mathbbm 1}/d$ on ${\bf C}^d$ is a
maximally entangled state given by 
\begin{equation}
|\Psi\rangle{} = \frac{1}{\sqrt{d}}\sum _{m=1}^{d}
|e_m\rangle{}\otimes|e'_m\rangle{}\ ,\ \ \ \ \{|e_m\rangle{}\}, \{|e'_m\rangle{}\}-\
{\rm basis\ in}\ {\bf C}^{d}\ .
\label{eq_entst}
\end{equation}

2) There is one to one correspondence between antiunitary matrices acting on ${\bf
C}^d$ and maximally entangled states in ${\bf C}^d\otimes{\bf C}^d$ which can be
expressed in terms of an arbitrary complete set of vectors ${\cal E}$ as
\begin{equation}
G \leftrightarrow |\Psi_G\rangle{} = \sqrt{d}\int_{\cal E} d\alpha |\alpha \rangle
\otimes |G\alpha \rangle \ .
\label{eq_defG}
\end{equation}

{\bf Proof} 

1) Follows directly from the Schmidt decomposition.

2) Take two complete sets ${\cal E}$ , ${\cal E}'$ and two antiunitary matrices
$G, G'$. Define $|\Psi_G\rangle{} $ by (\ref{eq_defG}) 
put $|\Psi_{G'}\rangle{} = \sqrt{d}\int_{\cal E'} d\beta |\beta\rangle{}\otimes
|G'\beta\rangle{}$ and compute the scalar product
\begin{equation}
\langle{}\Psi_{G'}|\Psi_G\rangle{} = d\int_{\cal E}d\alpha\int_{\cal E'}d\beta
\langle{}\beta |\alpha\rangle{} \langle{} G (G^{-1}G')\beta|G\alpha\rangle{} =
\int_{\cal E'} d\beta \langle{}\beta |(G^{-1}G')\beta\rangle{} \ .
\label{eq_scalpr}
\end{equation}
From (\ref{eq_scalpr}) it follows that $\|\Psi_{G'}\| =\|\Psi_G\| = 1$ and $\Psi_{G'}
= \Psi_{G}$ if and only if $G = G'$. Obviously, for any vector of the form
(\ref{eq_entst}) we can choose the unique antiunitary matrix satisfying $G e_j =
e'_j$.

\section{Entropy production as a measure of decoherence}
\label{sec_entprod}

For a generic quantum system $S$ interacting with an environment (e.g. measuring
apparatus) its initial pure state becomes mixed due to the increasing
system-environment entanglement.  Assuming that the reduced dynamics is given by a
completely positive map (\ref{eq_supero}), 
to measure the decoherence we may use the von Neumann entropy
\begin{equation}
E[{\bf X},\alpha] = S\bigl(\Phi_{\bf X}(|\alpha\rangle{}\langle{}\alpha|)\bigr) =
S(\sigma^{\alpha}[{\bf X}]) \ ,
\end{equation}
where $|\alpha\rangle{}\in {\cal H}_S$ is an initial pure state of the system and
$\sigma^{\alpha}[{\bf X}]$ is $k\times k$ correlation matrix with $(ij)$- element
$\langle{}\alpha | X^{\dagger}_j X_i | \alpha\rangle{}$. We extend this construction
to the case of discrete time  finite quantum dynamical system with the unitary
evolution $U$ interrupted by a measuring process (or generally interaction with an
environment) described by the partition of unity ${\bf X}$ or equivalently by the
%
\begin{equation}
E_t[{\bf X},U,\alpha] = S\bigl([\Phi_{U\bf
X}]^t(|\alpha\rangle{}\langle{}\alpha|)\bigr) = S(\sigma^{\alpha}[{\bf X}^t]) \ .
\label{eq_defet}
\end{equation}
The quantity of above, bounded by $\ln ({\rm dim}{\cal H}_S )$ can be strongly
dependent on the initial state of the system. 

Assume now that the system $S$ is finite i.e. ${\cal H}_S = {\bf C}^d$.  In order to
obtain a more universal measure we can average the entropy over a complete set ${\cal
E}$ of pure states $\{|\alpha\rangle{}\}$ . The entropy averaged with respect to
${\cal E}$  is equal to
\begin{equation}
E_t[{\bf X},U,{\cal E}]=\int_{\cal E} d\alpha E_t[{\bf X},U,\alpha]  \leq \ln {d} \ ,
\label{eq_et}
\end{equation}
and its increase (entropy production) characterizes the magnitude of the decoherence
process.

Since the entropy $E_t$ is bounded from above, its asymptotic production rate, $E_t/t$
tends to zero for $t\to \infty$.  On the other hand, we will be interested in the
initial production rate.  Studying a discrete dynamics we cannot define the derivative
$dE_t/dt$, but we may for instance study the entropy produced after each initial time
step.  Analyzing the trivial dynamics,  $U={\mathbbm 1}$, and measurement process
governed by projection operators, $X_j=P_j=(P_j)^2$, the entropy is produced only
once, and $E_t=E_1$ for all $t>0$. Therefore, to characterize in this situation the
unitary dynamics, and not the measurement process itself, we are going to use the
quantity $\Delta E=E_2-E_1$ as a measure of the  decoherence.  For comparison we
define the initial production of the partial ALF-entropy, $\Delta S=S_2-S_1$.

Defining the ALF entropy, which characterizes the unitary evolution $U$ one uses the
supremum (\ref{eq_qedyn}) over all operational partitions of unity. Let us emphasize
that there is no point in performing such a step by studying the initial decoherence
rate $\Delta E$.  Since the set of transformed operators,  $P_j\to P'_j= P_j V$
(${\mathbf P}=\{P_1,\dots ,P_k\}$) with
arbitrary unitary $V$ is also a valid identity resolution, then $E_t[{\bf P},U,{\cal
E}] = E_t[{\bf P'},VU,{\cal E}]$ so the supremum over all possible measurements will
be independent of the unitary dynamics $U$ studied.
 
It follows from Eq. (\ref{eq_st}) and Eq. (\ref{eq_Omega}) that the time dependent
entropy $S_t[{\bf X},U]$, which appears in the definition of the ALF-entropy and in
the semiclassical regime is related to the KS entropy, describes also the magnitude of
a certain decoherence process. However, this process involves maximally entangled state
of the system plus ancilla while the natural decoherence measure should be defined in
terms of the system alone, like $E_t[{\bf X},U,\alpha]$ or $ E_t[{\bf X},U,{\cal E}]$.

\medskip

In order to compare both entropies $E_t[{\bf X},U,{\cal E}]$ and  $S_t[{\bf X},U]$ we
need the following technical result.

Take the dynamical map $\Phi_{\bf Y}$ defined by the $k$-elements partition of unity
${\bf Y}= \{Y_1,Y_2,\dots ,Y_k\}$ as in (\ref{eq_supero}) and an arbitrary complete set
of vectors ${\cal E}$.  We use the notation $\sigma^{\alpha}[{\bf Y}]$ and $\sigma[{\bf
Y}]$ for the $k\times k$ correlation (density) matrices with matrix elements
$\langle{}\alpha |Y_j^{\dagger}Y_i |\alpha\rangle{} $ and $\frac{1}{d}{\rm
Tr}(Y_j^{\dagger}Y_i)$ respectively. We introduce also the tracial norm $\|A\|_1 =
{\rm Tr}\bigl[(A A^{\dagger})^{1/2}\bigr]$ and the Hilbert-Schmidt norm $\|A\|_2 =
\bigl[{\rm Tr}(A A^{\dagger})\bigr]^{1/2}$ for a matrix or an operator $A$ and the
entropy function $\eta (x) = -x\ln x$.

{\bf Theorem 2} 

\begin{equation}
A \geq S(\sigma[{\bf Y}]) - \int_{\cal E} d\alpha S(\sigma^{\alpha}[{\bf Y}]) \geq B
	\ ,
\label{theorem2}
\end{equation}

where 
\begin{equation}
A = \int_{\cal E} d\alpha \Bigl( \|\sigma[{\bf Y}]- \sigma^{\alpha}[{\bf Y}]\|_1
\ln k + \eta \bigl( \|\sigma[{\bf Y}]- \sigma^{\alpha}[{\bf Y}]\|_1
\bigr)\Bigr) \ ,
\label{eq_boundA}
\end{equation}
\begin{equation}
B = \frac{1}{2}\text{max} \Bigl\{\int_{\cal E} d\alpha \bigl( \|\sigma[{\bf Y}]-
\sigma^{\alpha}[{\bf Y}]\|_1\bigr)^2 , \int_{\cal E} d\alpha \bigl(\|\sigma[{\bf Y}]-
\sigma^{\alpha}[{\bf Y}]\|_2\bigr)^2 \Bigr\} \ .
\label{eq_boundB}
\end{equation}
{\bf Proof}. We use the inequalities for the relative entropy \cite{OP93,St95,AF01}
\begin{equation}
S(\rho |\omega) = {\rm Tr}\bigl( \rho\ln\rho - \rho\ln\omega\bigr) \geq \frac{1}{2}
{\rm max}\{\|\rho -\omega\|_1^2 , \|\rho -\omega\|_2^2\}\ .
\end{equation}
Putting $\rho = \sigma^{\alpha}[{\bf Y}]$ and $\omega = \sigma[{\bf Y}] =\int_{\cal
E}d\alpha \sigma^{\alpha}[{\bf Y}]$ and averaging over $d\alpha$ we obtain the lower
bound~(\ref{eq_boundB}).  The upper bound~(\ref{eq_boundA}) follows directly from the
Fannes inequality \cite{CNT87,AF01}, 
\begin{equation}
|S(\rho) - S(\omega)|\leq (\|\rho - \omega\|_1 \ln ({\rm
dim}{\cal H})+ \eta (\|\rho - \omega\|_1) \ .
\end{equation}
A basic consequence of the Theorem 2 is the inequality
\begin{equation}
2\ln d \geq S_t[{\bf X}\otimes {\mathbbm 1}_d,U] \geq E_t[{\bf X},U,{\cal E}] \leq \ln
d \ ,
\label{eq_mainineqaulity}
\end{equation}
and the fact that  $S_t[{\bf X},U] = E_t[{\bf X},U,{\cal E}] $ if and only if
$\sigma^{\alpha}[{\bf X}^t] = \sigma[{\bf X}^t]$ for almost all $\alpha$
(except of a set of measure zero). Numerical results show that for small times
both quantities are comparable $S_t[{\bf X},U] \simeq E_t[{\bf X},U,{\cal E}]$.
We can provide  some arguments in favour of this behavior in the case of the
complete set ${\cal P}_d $ of all pure states. 

Consider the fluctuations of the matrix elements of the $k\times k$ matrix
$\sigma^{\alpha}[{\bf Y}]$ treated as  random variables with respect of the uniform
measure over the set of all pure states ${\cal P}_d$. Deviation of a matrix element
from its average is given by
\begin{equation}
\delta = |\langle{}\xi|\sigma^{\alpha}[{\bf Y}]- \sigma[{\bf Y}]|\xi\rangle{}|\ ,
\label{eq_delta}
\end{equation}
where $|\xi\rangle$ is an arbitrary normalized vector from ${\bf C}^k$. Expectation
value of the operator $\sigma^{\alpha}[{\bf Y}]$ is equal
\begin{equation}
\langle{}\xi|\sigma^{\alpha}[{\bf Y}]|\xi\rangle{} =
\langle\alpha|\sum_{ij}\bar{\xi_j}\xi_iY^{\dagger}_jY_i|\alpha\rangle \ .
\end{equation}
The positive operator $B =\sum {\bar\xi}_i{\xi}_j Y^{\dagger}_iY_j < {\mathbbm 1}$,
i.e.  for any normalized vector $|\phi\rangle\in{\bf C}^d$, $\langle\phi |
B\phi\rangle \leq 1$.  So, it can be written in the form of convex sum of projectors
into its eigenvectors $|\Psi\rangle$, i.e. $B = \sum_l
b_l|\Psi_l\rangle\langle\Psi_l|$ ($0\leq b_l\leq 1$).  Let $\alpha_l$ denote
coefficients of the random normalized state $|\alpha\rangle$ with respect to the
eigenvectors of the operator $B$.  Finally
\begin{equation}
\delta = \left|\sum_l b_l (|\alpha_l|^2 - \frac{1}{d})\right| \ .
\end{equation}
The numbers $(|\alpha_l|^2 -d^{-1})$ take positive and negative values of the
order $d^{-1}$ but sum up to zero. However, when multiplied by another random
variables $b_l\in [0,1]$ they behave like ``steps of the random walk'' yielding a sum
of the order $\sqrt{d}\times d^{-1}$ and hence
\begin{equation}
\delta \lesssim \frac{1}{\sqrt{d}}\ .
\end{equation}
%
%
%
Therefore the fluctuations of the norm $\|\sigma[{\bf Y}]- \sigma^{\alpha}[{\bf
Y}]\|_1$ behaves like $k/\sqrt{d}$ .

In the time-dependent case it means that for $k^t \ll d$ we have
$S_t[{\bf X},U] \simeq E_t[{\bf X},U,{\cal E}]$. Moreover, a random
choice of $|\alpha\rangle{}\in {\cal P}_d$ gives typically $ S(\sigma^{\alpha}
[{\bf X}^t]) \simeq E_t[{\bf X},U,{\cal E}]$ .

\section{Decoherence in periodically measured baker map}
\label{sec_decoh}

To illustrate the results presented in previous section on a concrete example we
analyze the quantum baker map (\ref{quantmap}) subjected to periodic sequence of
measurement performed in the momentum basis.  Entire, non-unitary dynamics of the
system is described by the superoperator 
\begin{equation}
\rho'=\sum_{j=1}^k P_j^P U \rho U^{\dagger} P_j^P \ .
\label{bakmes}
\end{equation}
The set ${\bf P^P}$ of $k$ projection operators fulfills the identity resolution,
$\sum_j P_j^P={\mathbbm 1}$, since the measurement process corresponds to the partition
of phase space into $k$ equal intervals in momentum,
\begin{equation}
{\bf P} = \left\{ P_j^P: P_j^P= \sum_{i=(j-1)d/k+1}^{jd/k} | \tilde e_i \rangle
\langle \tilde e_i | \right\} ,
\end{equation}
where $\tilde e_i$ are the momentum eigenstates defined by (\ref{momeigv}) in ${\bf
C}^{d}$, and the size $d$ of the Hilbert space is an integer multiple of $k$.

\begin{figure}[hbt]
\begin{center} \includegraphics[width=10cm]{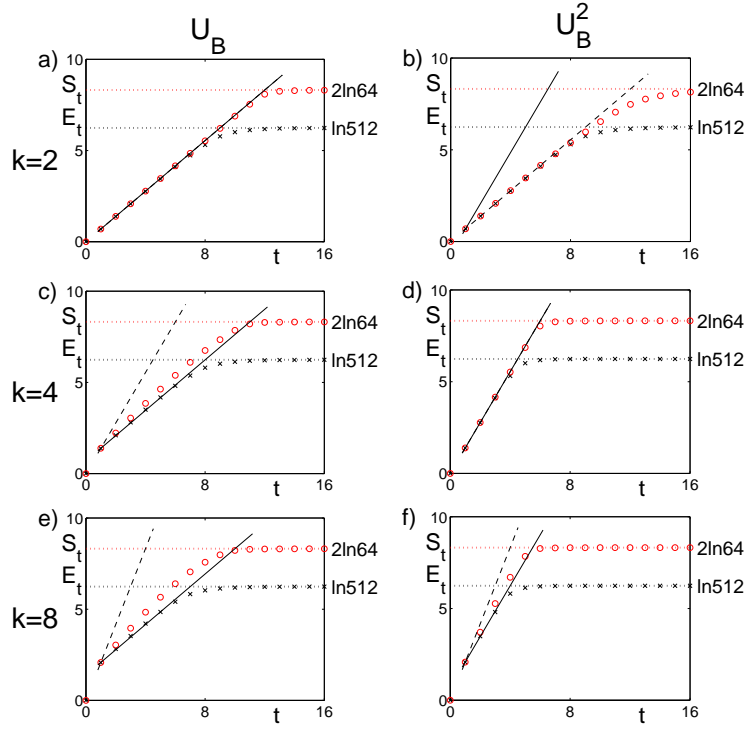} \end{center}
\caption{Initial growth of entropies $S_t[{\bf P^P},U]$ ($\circ$) and $E_t[{\bf
P^P},U,{\cal P}_d]$ ($\times$) computed for baker map $U=U_B$ (a), c) and e)) and its
square $U={U_B}^2$ (b), d) and f)) where partition ${\bf P^P}$ corresponds to division
of classical phase space into $k = 2 (\text{a) and b)}), k = 4 ((\text{c) and d)}), k
= 8 ((\text{e) and f)})$ equal intervals in momentum coordinate.}
\label{mompart}
\end{figure}

Iterating numerically quantum map (\ref{bakmes}) we compute how both entropies
$S_t[{\bf X},U]$ (\ref{eq_Omega}) and $E_t[{\bf X},U,{\cal E}]$ (\ref{eq_et}) vary in
time.  Fig.~\ref{mompart} presents the initial growth of the entropy $S_t[{\bf
P^P},U]$ and $E_t[{\bf P^P},U,{\cal P}_d]$, where we have averaged the entropy over
the entire set of pure states with the natural measure, ${\cal E} = {\cal P}_d$. As
the evolution operator $U$ we took the quantum baker map $U_B$ defined by
(\ref{quantmap}) part (a), (c) and (e), and its square ${U_B}^2$ (b), (d) and (f). The
partition ${\bf P^P}$ is composed of $k=2,4,8$ projection operators. Entropy $E_t$ is
averaged over a sample of $32$ randomly chosen pure initial states.  To guide the eye
we plotted solid lines corresponding to growth with the rate of classical KS~entropy,
which is equal $\ln 2$ in the case (a), (c) and (e) and $2\ln 2$ in (b), (d) and (f).
The slope of the dashed lines is equal to the maximal allowed growth of entropies,
equal to $\ln k$.  Aiming for the semicalssical regime, we have taken the maximal
dimensionality of the Hilbert space, which was allowed by the computer resources at
our disposal. In order to compute entropy $S_t$ one has to diagonalize matrices of
size $d^2$, so we could work with matrices size $d=64=2^6$.  To obtain the entropy
$E_t$ one needs to study the time evolution of density operators acting on ${\cal
H}_d$, so we succeed to work with systems of the size $d=512=2^9$. Entropy $E_t$
obtained for $d=64$ is smaller than $S_t$ according to the analytical bound
(\ref{eq_mainineqaulity}). As discussed in \cite{lozinski02} the size $d$ of the
Hilbert space determines only the saturation level ($E_t(t\rightarrow\infty)$), but
does not influences the initial entropy rate.  If the measurement scheme is tuned to
the classical dynamics, i.e. case (a) and (d), the rate of the initial growth of both
entropies coincides with the classical dynamical entropy $h[T]$ of the map which is
equal $\ln 2$ for the baker map, and $2\ln 2$ for its square. If the resolution of the
measurement is not sufficient --- (b) the classical chaos cannot fully manifest itself
and $\Delta E$ and $\Delta S$ are smaller than $ h[T]$ and equal to $\ln k$.  In the
opposite case, panel (c), (e) and (f),  a finer resolution of the measurement ($\ln k$
> $h[T]$) allows  for the decoherence with the rate faster than it can be expected
from the classical entropy.  Hence such a measurement can be responsible for entropy
production faster then it may be predicted basing on the degree of the classical
chaos. As visible in panel (e) this effect is larger for the entropy $S_t$.

\begin{figure}[hbt]
\begin{center} \includegraphics[width=10cm]{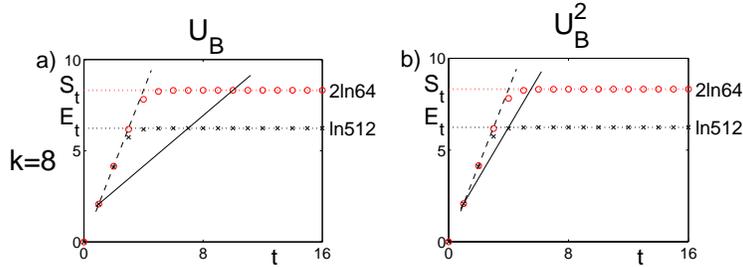} \end{center}
\caption{As in the Fig. \ref{mompart} (e) and f) with $k = 8$ for the measurement in a
random basis ${\bf P^R}$.}
\label{ranpart}
\end{figure}

To demonstrate other features of the measurement process
 we investigated the time dependence of both
entropies $S_t[{\bf P^R},U]$ and $E_t[{\bf P^R},U,{\cal P}_d]$ for different choices of
the partitions ${\bf P^R}$. Fig.~\ref{ranpart} shows the initial growth of both
entropies calculated with the same evolution operators as in Fig.~\ref{mompart}.
However, the partition ${\bf P}^R$ was obtained by rotating the projective partition
${\bf P^P}$ by a random unitary matrix $V$, namely ${P_j^R} = V{P_j^P}V^\dagger$, for
all $j=1,\dots ,k$. The label ${\bf ^R}$ decorating the symbol ${\bf P^R}$ of the
partition emphasizes the fact that the measurement is performed in a random basis.
Such a measurement will give $k$ different results with equal probabilities
$\text{tr}(P_j^R) /d = 1/k$.  In Fig.~\ref{ranpart} both entropies initially increase
with a nearly maximal slope which is equal to $\ln k = \ln 8$ (dashed line). Unitary
evolution operator $U$ does not influence the behavior of both entropies, and the data
presented in both plots (a) and (b) hardly differ.

The fact that the rotated partitions leads to an almost maximal allowed growth of both
entropies may be explained by the following argument.  Both $S_t[{\bf X},U]$ and
$E_t[{\bf X},U,{\cal P}_d]$ may be expressed as the von Neumann entropy of a mixed state
obtained by the operator $[\Phi_{U\bf X}]^t$ applied to a pure state (see Eqs.
(\ref{eq_Omega}) and (\ref{eq_defet})). Let us denote the rotated partition by ${\bf
Y} = V{\bf X}V^\dagger$, where $V$ is an arbitrary unitary matrix. From definition of
the operator $\Phi_{\bf X}$ (\ref{eq_supero}) we have
\begin{equation}
[\Phi_{U\bf Y}]^t(\rho) = [\Phi_{UV{\bf X}V^\dagger}]^t(\rho)
= V[\Phi_{V^\dagger UV\bf X}]^t(V^\dagger \rho V)V^\dagger
= V[\Phi_{U'\bf X}]^t(\rho')V^\dagger,
\end{equation}
where the initial state $\rho$ becomes rotated, $\rho'=V^\dagger \rho V$, and the
evolution $U$ is replaced by $U' = V^\dagger U V$. The von Neumann entropy depends
only on the spectrum of density matrix, so $S([\Phi_{U\bf Y}]^t(\rho)) = S([\Phi_{U'\bf
X}]^t(\rho')$. In the case of the entropy $S_t$ the evolution operator is a tensor
product $U = U_S \otimes U_A$, so are the partition ${\bf X} = {\bf X}_S \otimes
{\mathbbm 1}$, and the matrix $V = V_S \otimes {\mathbbm 1}$. Since we choose randomly
only $V_S$, the state of the ancilla is not important. When $\rho$ is a maximally
entangled state so is $\rho'$, since those state are invariant under local operations.
Hence we obtain $S_t[{\bf Y},U] = S_t[{\bf X},U']$. The case of the entropy $E_t$ is
simple, since here the averaging over all initial pure states $| \alpha \rangle$
automatically cancels the difference between $\rho$ and $\rho'$. Therefore the
equality $E_t[{\bf Y},U,{\cal P}_d] = E_t[{\bf X},U',{\cal P}_d]$ holds. As we can see in
both cases the randomly rotated partition ${\bf Y}$ is equivalent to model with
original partition ${\bf X}$ and the evolution operator rotated into $U'=V^\dagger U
V$, where $V$ is a random unitary matrix, generated according to the Haar measure on
$U(N)$.  Although the spectra of $U$ and $U'$ are equal, the eigenvectors of $U'$ are
random,  and the operator $\Phi_{U'\bf X}$ generates the maximal growth of the von
Neumann entropy $\Delta E\approx \min \{ \ln d,\ln k\}$. Here $\ln d$ represents the
maximal von Neumann entropy of the mixed state, while the measurement
 $\bf X$ transforms any pure state into a mixture with entropy bounded by $\ln k$.

This argument shows that for finite systems taking the supremum over all
possible partitions of unity leads to the maximal allowed growth of both
entropies irrespectively of the analyzed unitary dynamics $U$.  Note that the
upper bound $\ln d$ is only slightly larger that the average quantum dynamical
entropy of a random unitary matrix $U$ distributed according to the Haar
measure on $U(d)$ \cite{SZ99}. 

\begin{figure}[hbt]
\begin{center} \includegraphics[width=10cm,angle=-90]{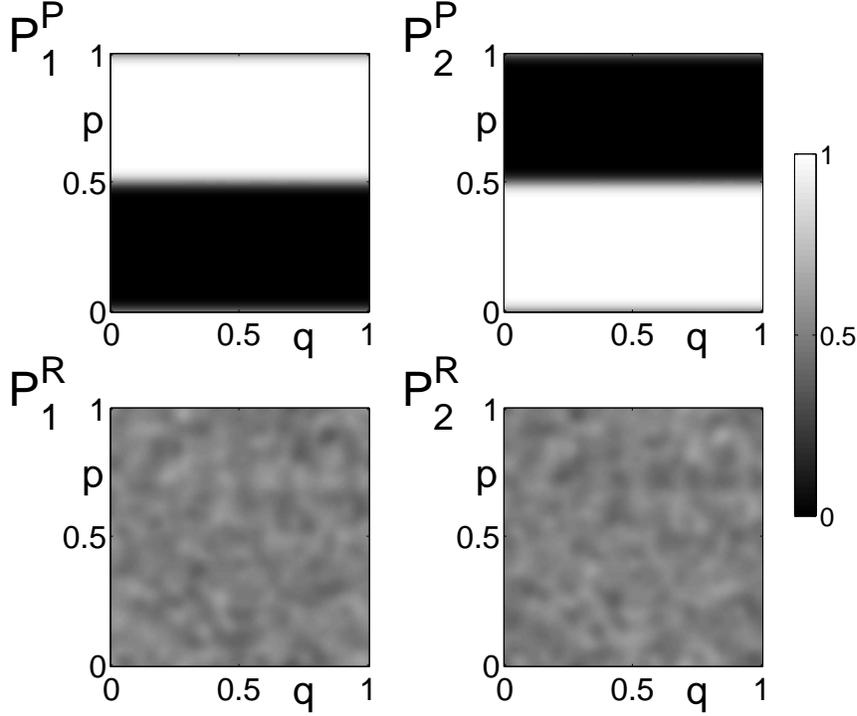} \end{center}
\caption{Phase-space representations of partitions 
${\bf P^P} = \{P_1^P, P_2^P\}$ 
 and ${\bf P^R} = \{P^R_1, P^R_2\}$ consisting of $k=2$ operators. The
    real function $x_j(q,p) = \langle q,p | X_j^\dagger X_j | q,p
    \rangle$ is plotted in dark scale, where $X_j$ denotes one of
      $P_1^P, P_2^P, P^R_1, P^R_2$. 
    The partition ${\bf P^P}$ corresponds to partition on
    upper and lower half in momentum coordinate while ${\bf P^R}$ is
    partition corresponding to the projections in a
     a randomly selected basis.  Note that the coherent states
    representations of $P^R_1$ and $P^R_2$ are delocalized in 
     the phase space.}
\label{partrep}
\end{figure}

On the other hand, one may pose a question, how to restrict the set of possible
partitions, such that the rate of growth of von Neumann entropy could
correspond to the dynamical entropy of the classical system. To analyze this
problem compare the properties of the momentum partition ${\bf P^P}$ and the
random partition ${\bf P^R}$ in the phase space. To represent the partition
member we make use of the Husimi-like representation,
\begin{equation}
x_j(q,p) \equiv  \langle q,p | X_j^\dagger X_j | q,p \rangle.
\end{equation}
Here $| q,p \rangle$ denotes the Gaussian states localized on torus,
the same as in Ref.~\cite{vallejos99,lozinski02}. In Fig.~\ref{partrep} we show the
phase-space representation of two partitions of unity ${\bf P^P} = \{P_1, P_2\}$, and
${\bf P^R} = \{VP_1V^{\dagger}, VP_2V^{\dagger}\}$, each consisting of $k=2$
operators. ${\bf P^P}$ denotes the partition into equal intervals in momentum
coordinates and ${\bf P^R}$ is a partition into two subspaces of equal size determined
by a random unitary matrix $V$. As may be seen in the picture, the operators $P^P_1$
and $P^P_2$ are by construction localized in lower (upper) region of the phase space,
while $P^R_1$ and $P^R_2$ are totally delocalized.

These results suggest that in order to predict the decoherence rate $\Delta E$ (and
$\Delta S$) one should consider only these partitions ${\bf P^C}$, which have a
well-defined classical limit. This is the case if each of the operator $X_j$ is
localized on a subset $\epsilon_j \subset \Gamma$ of phase space, so in the classical
limit $X_j(q,p) \rightarrow \chi_{\epsilon_j}(q,p)$, where $\chi_{\epsilon_j}$ denotes
the characteristic function of $\epsilon_j$.

The discussion of above has some important consequences for the decoherence processes
in quantum systems having chaotic classical limits with the K-S entropy $h[T]>0$.
Namely, if only the interaction with the environment can be described by a partition
of unity having a well defined classical limit (and $\ln k \geq h[T]$) we expect that
the decoherence effects give the entropy production per single time step of the order
$h[T]$ for the generic initial conditions. The linear entropy increase has to break
down for times $t$ of the order of $\ln d / h[T]$~\cite{zurek94}.  On the contrary, if
the partition of unity is chosen randomly and has no well-defined classical limit we
do not expect any restrictions on entropy production rate, except the general upper
bound $\Delta S \leq \ln k$, related to the number $k$ of the Kraus operators.  

\section{Occurrence of free-independent  variables}
\label{sec_free_indep}

Semiclassical arguments of the previous sections do not explain the striking
phenomenon observed in numerical computations of $S_t[{\bf X},U]$ and $E_t[{\bf
X},U,{\cal P}_d]$ as presented in Fig.~\ref{mompart} (a) and (d) and Fig.~\ref{ranpart}.
Namely for the two situations:

a) the quantum system with chaotic classical limit and the semiclassical, projection
valued choice of the partition ${\bf P} = \{P_1, P_2,\dots ,P_k\}, {\rm tr} P_j = d/k$
satisfying $\ln k \leq h[T]$,

b) nontrivial $U$ and the random choice of the partition ${\bf P}$.

\medskip

$S_t[{\bf X},U]$ grows almost exactly linearly like $t\ln k$ and then rapidly
saturates at the  maximal value $2\ln d$. The entropy $E_t[{\bf X},U,\alpha]$ with a
random choice of $|\alpha\rangle{}$ follows the same plot up to its maximal value $\ln
d$. This means that in both cases a) and b) the correlation density matrices possess a
very special structure corresponding to the maximal admissible entropy, 
\begin{equation}
\sigma[{\bf P}^t]_{i_1,\dots ,i_t;j_1,\dots ,j_t}   
=d^{-1} {\rm tr} \bigl(P_{j_1}U^{\dagger} P_{j_2}U^{\dagger}\dots P_{j_t}
 P_{i_t}\dots U P_{i_2}U P_{i_1}\bigr) \simeq \frac{1}{k^t}
 \delta_{i_1j_1}\cdots\delta_{i_tj_t} \ ,
\label{eq_corrdenmat}
\end{equation}
and similarly for a typical vector $|\alpha\rangle$,
\begin{equation}
\sigma^{\alpha}[{\bf P}^t]_{i_1,\dots ,i_t;j_1,\dots ,j_t}   
= \langle\alpha | \bigl(P_{j_1}U^{\dagger} P_{j_2}U^{\dagger}\dots P_{j_t}
 P_{i_t}\dots U P_{i_2}U P_{i_1}\bigr)| \alpha\rangle \simeq \frac{1}{k^t}
 \delta_{i_1j_1}\cdots\delta_{i_tj_t}\ ,
\label{eq_corrdenmat1}
\end{equation}
under the condition of $ k^t \ll d$.
\par
The simple form of the correlation functions 
(\ref{eq_corrdenmat}) and (\ref{eq_corrdenmat1}) suggests the existence of a certain
statistical law satisfied by the noncommutative variables $\{P_j, U\}$
with respect to the tracial state or a typical pure state $|\alpha\rangle$. This law
should be strictly obeyed in the limit $d\to\infty$ but even for relatively low
dimensions reproduces the data with very good accuracy. Such situations are common in
Nature. Gaussian and Poisson probability distribution are very successful in
describing experimental data while their rigorous derivations involve limit theorems
with strong statistical independence assumptions.

We advance  the following statistical hypothesis:

{\em For both cases a) and b) and large Hilbert space dimensions $d$ the operators
$\{A , U\}$ behave asymptotically like free-independent random variables with respect
to the tracial state or a typical pure state.}

Here $A$ is an arbitrary observable with the spectral measure $(P_1, P_2,\dots P_k)$. 

We have to explain now the notion of {\em free-independence}. In the classical
probability theory (complex) random variables form a commutative $*$-algebra and the
probability measure defines a positive normalized functional $f\mapsto
\langle{}f\rangle{}$ - the {\em average value}. The random variables $f_1,
f_2,\dots ,f_n$ are called statistically independent if
\begin{equation}
\langle{}f_1 f_2 \dots  f_n\rangle{} = \langle{}f_1\rangle{}\langle{}f_2\rangle{}\cdots
\langle{}f_n\rangle{}\ .
\label{eq_statindep}
\end{equation}
In {\em noncommutative probability} the basic object is a unital generally
noncommutative $*$-algebra ${\cal A}$ with a state (positive normalized functional)
$\phi$. Due to the noncommutativity the average $\phi (x_1 x_2\cdots x_m); x_j\in{\cal
A}$ depends on the order of random variables. Hence, the direct extension of the
definition (\ref{eq_statindep}) is not very interesting and essentially corresponds to
product states on tensor product algebras. Instead, in noncommutative probability we
have different notions of independence which take into account possible algebraic
relations between random variables (e.g. CCR, CAR, etc \cite{RB97}). In the last
decade the so-called free families of random variables (or free-independence)
introduced by Voiculescu \cite{Voilculescu91,VDN92} attracted attention of physicists
mainly due to the relations with random matrices theory. 

Denote by $w(x)$ an arbitrary polynomial in variables $x, x^{\dagger}\in {\cal A}$.
The collection of noncommutative random variables $x_1, x_2,\dots ,x_k$ is called {\em
free-independent} if
\begin{equation}
\phi (w_1(x_{p(1)})w_2(x_{p(2)})\cdots w_m(x_{p(m)})= 0 \ ,
\end{equation}
whenever $\phi (w_j(x_{p(j)})=0$ and $p(j)\ne p(j+1)$ for all $j=1,2,\dots ,m$, $p(j)\in
\{1,2,\dots ,k\}$.

It has been proved that the Wigner semicircular probability distribution is a
consequence of the central limit theorem for free-independent random variables
similarly to the origin of Gaussian probability distribution in the context of
statistically independent commutative variables. Moreover, the free-independent
variables naturally arise as limits of large random matrices
\cite{Voilculescu91,VDN92}. The consequences of free independence are illustrated by
the following example.

{\bf Example}. Take a family of orthogonal projections $P_1,P_2,\dots ,P_k$ and a unitary
$U$, all from the algebra ${\cal A}$ with the state $\phi$. Assume that for any $A=
\sum a_j Pj$ the pair of random variables $\{A, U\}$ is free-independent and moreover
\begin{equation}
\phi (P_j)= \frac{1}{k}\ ,\ j=1,2,\dots ,k\ ,\ \ \phi (U^n) = \phi({U^{\dagger}}^n) = 0\ ,\
n=1,2,\dots  \quad \ .
\label{eq_starfree}
\end{equation}
Then
\begin{equation}
\phi \bigl(P_{j_1}U^{\dagger}P_{j_2}U^{\dagger}\cdots P_{j_n}U^{\dagger}UP_{i_m}\cdots
UP_{i_2}UP_{i_1}\bigr) = \delta_{nm} \frac{1}{k^n}\delta_{i_1j_1}\cdots
\delta_{i_nj_n}\ .
\label{eq_phi}
\end{equation}
{\bf Proof}. Put $Q_j = P_j - 1/k$, then 
\begin{equation}
Q_iQ_j = \delta_{ij}Q_i - k^{-1} (Q_i +Q_j) + \delta_{ij} k^{-1} - k^{-2}\ ,\ 
\phi(Q_j) = 0\ .
\end{equation}
Hence the LHS of (\ref{eq_phi}) is a linear combination of the terms of the form 
\begin{equation}
\phi \bigl( {U^{\dagger}}^{n_1} Q_{k_1}{U^{\dagger}}^{n_2}\cdots U^{m_1}
Q_{k_1}U^{m_2}\cdots\bigr) \ ,
\end{equation}
which due to the freeness and Eqs. (\ref{eq_starfree}, \ref{eq_phi}) are all equal to
zero except the terms which do not contain nontrivial powers of $U$ and $U^{\dagger}$.
This can happen, however, for $n=m$ only. In this case we can easily prove
(\ref{eq_phi}) by induction.

The relation (\ref{eq_phi}) corresponds to the phenomenon observed in the numerical
computations of $S_t[{\bf X},U]$ and $E_t[{\bf X},U,{\cal P}_d]$ and described by
(\ref{eq_corrdenmat}) and (\ref{eq_corrdenmat1}). This justifies our hypothesis
formulated above.  In the case b) this hypothesis is not surprising due to the random
choice of the partition and the generic relations between free random variables and
random matrices.  Similarly, we would expect the same phenomenon for the fixed
partition and the random choice of the unitary matrix. On the other hand, for the case
a) it seems to be a new characterization of chaotic quantum systems in terms of
"quantum-probabilistic" relations between the dynamics and the measurement
(coarse-graining) procedure. 

\section{Concluding remarks}
\label{sec_conclud}

We have analyzed the decoherence in an open quantum system,
the classical analogue of which is chaotic.  The decoherence may be quantified
by the rate of increase of the von Neumann entropy of the initially pure
states.  We have found an explicit upper bound for the rate of the von Neumann
entropy given by the partial entropy used to define the ALF dynamical entropy.
The later quantity is related to the Kolmogorov--Sinai entropy of the
corresponding classical system.  Hence our findings allow us to established a
further relation between the speed of decoherence in open quantum systems and
the degree of classical chaos.

Such a relation, demonstrated in several earlier works \cite{zurek94,MP00,BPS02}
holds if some additional assumptions concerning the coupling of the system
investigated with an environment (the measurement process) are made. In
particular, we proposed to study the scheme of {\sl random measurements}, in
which the usual Kraus operators $X_i$, which represent projectors on some well
defined fragments of classical phase space, are replaced by operators obtained
by random matrices, $X'_i=V X_i V^{\dagger}$.  In such a case the rate of von
Neumann entropy becomes maximal (with probability one, with respect to the
choice of random matrix $V$).  Thus the decoherence depends only on the kind of
the measurement performed (the number of the Kraus operators or the
dimensionality of the system), and is independent of the quantum unitary
dynamics $U$, and of the degree of chaos (Lyapunov exponent, KS dynamical
entropy) of the corresponding classical system. 

From a practical point of view it is therefore natural to ask, for which class
of measurement procedures the relation between classical chaos and the degree
of quantum decoherence is still valid.  Although we are not in position to
formulate mathematically rigorous sufficient conditions, which would imply such
a relation, our numerical evidence allows us to advance the following
conjecture. The interaction with an environment induces decoherence related to
the degree of the classical chaos, if the measurement (Kraus) operators have a
well defined  classical limit. In other words, the   coherent states
representation of each of Kraus operators needs to be well localized in certain
fragments of the classical phase space.

More formally, the maximal entropy growth, and its independence of the unitary
dynamics,  may be analytically derived from an assumption that unitary operator
$U$ and an arbitrary combination of the projector operators $P_i$ are
free independent.  Obviously this statement is of a statistical nature, and
does not allow one to draw rigorous conclusion for a concrete set of projection
and evolution operators.  The free random variables approach concerns entire
ensembles of operators and enables us to formulate exact statements concerning
the decoherence rate in the limit of large Hilbert space dimension.
Nevertheless, for practical purposes one may choose a set of arbitrary test
states $\phi$ and check whether property (\ref{eq_phi}) is approximately fulfilled
for the analyzed unitary map $U$ and measurement ${\bf P}$. It is worth to
emphasize that the free--independence condition can be formulated as a
condition for a pair of genuinely quantum objects -- an observable and an
unitary quantum map -- without any reference to the classical notions.
Therefore the idea of quantum chaos may be extended to systems without obvious
classical counterparts or to dynamics which do not satisfy standard assumptions
concerning the spectral fluctuations. 

It is a pleasure to thank J.~P.~Keating, J.~Marklof, W.~S{\l}omczy{\'n}ski and
G.~Tanner for fruitful discussions.  We are grateful to P.  Garbaczewski for
organizing the Winter School "Quantum Dynamical Semigroups" in Lądek in February 2002,
where this work has been initiated.  Financial support by Komitet Badań Naukowych is
gratefully acknowledged.

\end{document}